\documentclass[nofootinbib,aps,a4paper,10pt,letterpaper,superscriptaddress,onecolumn,showkeys,times,eqsecnum]{revtex4-2}
\usepackage[utf8]{inputenc}
\usepackage[T1]{fontenc}
\usepackage{cmap}
\usepackage{color}
\usepackage{epsfig}
\usepackage{amsmath}
\usepackage{graphicx}
\usepackage{amsfonts}
\usepackage{dsfont}
\usepackage{graphicx}
\usepackage{dcolumn}
\usepackage{bm}
\usepackage{epstopdf}
\usepackage{float}
\usepackage{subcaption}
\usepackage{pbox}
\usepackage{array}
\usepackage{amssymb}

\usepackage{hyperref}   
\definecolor{oxfordblue}{rgb}{0.0, 0.13, 0.28}
\definecolor{burgundy}{rgb}{0.5, 0.0, 0.13}
\definecolor{darkolivegreen}{rgb}{0.33, 0.42, 0.18}
\definecolor{darkblue}{rgb}{0,0,0.5}
\definecolor{richcarmine}{rgb}{0.84, 0.0, 0.25}
\definecolor{darkblue}{rgb}{0,0,0.5}
\definecolor{bluer}{rgb}{0.00,0.50,0.75}{}
\hypersetup{colorlinks=true, citecolor=blue, linkcolor=magenta,
	urlcolor = darkblue, filecolor=magenta}
\begin{document}
	\title{Dimensional Phenomenology in Polymeric Quantization Framework}
	
\author{\textbf{ Kourosh Nozari}}
\email{Knozari@umz.ac.ir}
\affiliation{Department of Physics,
	Faculty of Basic Sciences, University of Mazandaran,\\\small {\it P.O. Box 47416-95447,Babolsar, Iran}}

\author{\textbf{Hamed Ramezani}}
\email{h.ramezani01@umail.umz.ac.ir}
\affiliation{Department of Physics,
Faculty of Basic Sciences, University of Mazandaran,\\\small {\it P.O. Box 47416-95447,Babolsar, Iran}}
	
\date{\today}
\begin{abstract}
  In this paper, we study the statistical mechanics within the polymer quantization framework in the semiclassical regime. We apply a non-canonical transformation to the phase space variables. Then, we use this non-canonical transformation to calculate the deformed density of states of the $2n$-dimensional phase space, which encompasses all polymer effects. In the next step, some thermodynamic features of a system of $n$-dimensional harmonic oscillators are studied by computing the deformed partition function. The results show that the number of microstates decreases because there is an upper bound on the momentum within the polymer framework. We found that in the high-temperature regime, when the thermal de Broglie wavelength is close to the Planck length, $n$ degrees of freedom of the system are frozen in this setup. In other words, there is an effective reduction in space dimensions from $n$ to $\frac{n}{2}$ in the polymeric framework, which also signals the fractional dimension for odd-dimensional oscillators.

\vspace{0.5cm}
\end{abstract}	

	\maketitle
	\section{Introduction}
In the phenomenological theories of quantum gravity, there is multiple pieces of evidence that the structure of spacetime at very small length scales is not continuous and smooth but rather has a lattice and discontinuous structure. This phenomenon, caused by the effects of quantum gravity at the Planck scale, is a complex dynamical phenomenon sometimes referred to as Wheeler spacetime foam \cite{Wheeler:1957mu,Gibbons:1976ue,HI-1979}. These concepts have been proposed in theories as diverse as string theory, loop quantum gravity (LQG), and doubly special relativity (DSR), and the existence of a minimal measurable length, preferably of the order of the Planck length $(l_{_{\rm Pl}}= \sqrt{\hbar G/{c^3}\;}\approx 1.6\times 10^{-35}\,m)$, is one of the common consequences and a central feature in these frameworks \cite{String1,String2,String3,String4,LQG1,LQG2,LQG3,LQG4,DSR1,DSR2,DSR3,DSR4,DSR5}.

In the Heisenberg uncertainty principle (HUP) framework, the measurement precision of a test particle's position is not fundamentally restricted, allowing the minimal uncertainty in position measurement, $\Delta{\rm q_0}$, to zero. The generalized uncertainty principle (GUP) is a modification of the Heisenberg uncertainty principle. GUP models and DSR are related to string theory and loop quantum gravity, suggesting modified dispersion relations. These phenomenological aspects of quantum gravity suggest the existence of a minimal measurable length and a maximal momentum or energy for a test particle \cite{QGML1,QGML2,GUPUV1,GUPUV2,GUPUV4}.

Among the models that deal with the idea of the existence of a minimal measurable length scale, the so-called polymer quantization of a dynamical system uses a method similar to the effective models of $\rm {LQG}$. It is a somewhat unconventional quantum representation for the canonical commutation relations of quantum mechanical systems \cite{L95,L97,L03,Date:2006cx}. The minimal length scale, which here is known as the polymer length scale, is encoded in the Hamiltonian of the system. Thus, instead of a deformed algebraic structure coming from the noncommutative phase space variables, deformation shows itself in the Hamiltonian function, which is known as {\textsl{\it{polymerization}}} \cite{corichi,corichi2,PLMRUR1,PLMRUR2}. This minimal length has profound physical consequences in statistical mechanics and thermodynamics \cite{NG14,Ch-A}. The path integral representation of a polymer quantized scalar field was studied in \cite{Kajuri:2014kva}.

These phenomenological models that we mentioned above are a flat limit of the quantum gravity proposal and show the deformation of the density of states at the high-temperature regime, which typically leads to the self-exclusion of microstates. These properties lead to a reduction in the degrees of freedom and, subsequently, a reduction in the dimensions of the space for the desired system~\cite{Fityo,NHG15,NR17,UV-Reduction1,UV-Reduction2,Ramezani:2024stm}.

In this work,  we examine thermodynamic dimensional reduction via the statistical mechanics of n-dimensional harmonic oscillators in the semiclassical regime, using the polymer quantization model. By computing the deformed partition function, we show that the number of microstates reduces significantly in the high-temperature regime, effectively reducing space dimensions at high energies. This model also exhibits features reminiscent of fractional dimensions~\cite{Frac1,Frac2} for odd-dimensional harmonic oscillator systems.

The structure of the paper is as follows: In Section 2, the statistical mechanics in the polymer quantization framework is introduced. The standard canonical representation of the polymer framework is briefly presented, and its non-canonical chart is investigated. The polymeric density of states of 2n-dimensional phase space, in non-canonical form, is formulated, and then the corresponding partition function is derived.
In section 3, the thermodynamics of a system of $n$-dimensional harmonic oscillators is studied within the framework of polymer quantization. The results for $2D$ and $3D$ harmonic oscillators are presented in detail. Section 4 is devoted to the summary and conclusions.

\section{Polymeric statistical mechanics}

\subsection{Polymerization}
In the formulation of classical mechanics, the Hamiltonian is based on the Poisson algebra $\{\mathbf {q_i},\mathbf {p_j}\}=\delta_{ij}$ where $(\mathbf q,\mathbf p)$, are phase space canonical variables and define the kinematics of the classical system. In quantum mechanics, this classical Poisson algebra is substituted by its quantum counterpart, the Heisenberg algebra $[\hat{\mathbf q}_i,\hat{\mathbf p}_j]=i\hbar\, \hat{\delta}_{ij}$. This quantum algebra leads to the Heisenberg uncertainty principle. The standard Schr\"{o}dinger picture represents operators on the Hilbert space ${\mathcal H}={\mathcal L}^2({\mathbb R},d\mathbf q)$, the space of square-integrable functions with respect to the Lebesgue measure $d\mathbf q$ on the real line ${\mathbb R}$. While the Stone-Von Neumann theorem implies a unique representation of the commutation relation, the polymer representation offers a nontrivial alternative that supports the existence of a minimal measurable length scale called the polymer length scale \cite{corichi} and is examined in the symmetric part of the {\rm LQG} \cite{L03,PQ3}. This representation is based on the Hilbert space ${\mathcal H}_{\rm \mathbf p}={\mathcal L}^2({\mathbb{R} }_{_d},d\mathbf q_{_H})$, where the Lebesgue measure $d\mathbf q$ and the real line ${\mathbb R}$ are replaced by the Haar measure $d\mathbf q_{_H}$ and the real line ${\mathbb R}_{_d}$ with the discrete topology, respectively \cite{PQ4}. There is a dimensional parameter $\beta$ that effectively describes the additional structure in the polymer representation and the ordinary Schr\"{o}dinger picture is recovered in the continuum limit, $\beta\rightarrow 0$ \cite{corichi}. In the $\hbar\rightarrow\,0$ limit, this yields an effective, ${\beta}$-dependent classical theory interpretable as a classical discrete theory. It is also worth mentioning that this effective theory can be obtained directly from standard classical theory using the Weyl operator, without relying on the polymeric representation \cite{corichi2}. Polymeric effects, encoded in the deformed Hamiltonian, are covered in the density of states.

In polymer quantum mechanics, the position coordinate $\mathbf q$ is discretized with parameter $\beta$, leading to the absence of momentum operator $\hat{\mathbf p}$ as its displacement generator~\cite{L03,PQ3}. However, the Weyl exponential operator, analogous to the momentum operator in continuous space, plays a similar role in discrete space. The algebra generated by the exponentiated forms of $\mathbf q$ and $\mathbf p$ is given by\cite{corichi,BarberoG:2013epp}
\begin{equation}\label{WA1}
U(\alpha)=e^{\frac{i\alpha \hat{\mathbf q}}{\hbar}}\,\,\,\,\,;\,\,\,\,\,V(\beta)=e^{\frac{i\beta \hat{\mathbf p}}{\hbar}}
\end{equation}
where $\alpha$ and $\beta$ have dimensions of momentum and length, respectively, the Canonical Commutation Relations become,
\begin{equation}\label{WA2}
U(\alpha).V(\beta)=e^{-\frac{i\alpha\beta}{\hbar}}V(\beta).U(\alpha),
\end{equation}
and
\begin{equation}\label{WA3}
U(\alpha_1).U(\alpha_2)=U(\alpha_1+\alpha_2)\,\,\,;\,\,\,V(\beta_1).V(\beta_2)
=V(\beta_1+\beta_2).
\end{equation}

 The derivative of $f(\mathbf q)$ with respect to the discrete $\mathbf q$ can be approximated using the Weyl operator as follows\cite{corichi2},

\begin{eqnarray}\label{FWD}
{\partial_{\mathbf q}f(\mathbf q)\approx\frac{1}{2{\beta}}[f({\beta}-\mathbf q)
+f(\mathbf q+{\beta})]}=\frac{f(\mathbf q)}{2{\beta}}\Big(-\widehat{e^{-i\mathbf p{\beta}}}
+\widehat{e^{i\mathbf p{\beta}}}\Big)=\frac{if(\mathbf q)}{{\beta}}\widehat{\sin({\beta} \mathbf p)},
\end{eqnarray}
Differentiating the above equation, the following equation is obtained,
\begin{eqnarray}\label{SWD}
\partial_{\mathbf q}^2f(\mathbf q)\approx\frac{1}{{\beta}^2}\Big[f(\mathbf q+{\beta})-\big(f({\beta}-\mathbf q)+2
f(\mathbf q)\big)\Big]=\frac{-2f(\mathbf q)}{{\beta}^2}
\Big(\widehat{1-\cos({\beta} \mathbf p)}\Big).
\end{eqnarray}
Inferring from the relations (\ref{FWD}) and (\ref{SWD}), the polymeric process for the finite value of ${\beta}$ is defined as follows
\begin{eqnarray}\label{Polymerization}
\hat{\mathcal P}\rightarrow\,\frac{\widehat{\sin({\beta} \mathbf p)}}{{\beta}},
\hspace{1cm}\hat{\mathcal P}^2\rightarrow\,\frac{2\Big(1-
\widehat{\cos({\beta}\mathbf  p)}\Big)}{{\beta}^2}.
\end{eqnarray}
Therefore, from the above relation (\ref{Polymerization}), the polymer transformation ${\mathcal {P}}[F]$  can be defined as follows\cite{corichi2},
\begin{eqnarray}\label{PT}
{\mathcal{P}} [\mathbf p]=\frac{\sin({\beta} \mathbf p)}{{\beta}},\hspace{1.5cm}{\mathcal{P}}[\mathbf p^2]=
\frac{2\Big(1-\cos({\beta} \mathbf p)\Big)}{{\beta}^2},\hspace{1.5cm}{\mathcal{P}}[F(\mathbf q)]=F(\mathbf q).
\end{eqnarray}

Similarly, one can find polymer transformations of higher powers of momentum $\mathbf p$. A classical system is {\it polymerized} when transformation \eqref{PT} is applied to its Hamiltonian formalism.

 As it's clear from the polymerized transformation~(\ref{Polymerization}), the momentum is periodic and bounded: $\mathbf p\in[-\frac{\pi}{{\beta}},+\frac{\pi}{{\beta}})$, and in the limit of ${\beta}\rightarrow\,0$, this polymerized transformation reduces to its classical counterpart, and the canonical momentum returns to its ordinary interval, $\mathbf p\in(-\infty,+\infty)$. So, in the polymer phase space, the momentum topology is $S^1$, unlike the classical case, which is $\mathbb{R}$. This nontrivial topology leads to unusual thermodynamic results, such as a reduction in the number of degrees of freedom, as we will examine in the next subsection.

 \subsection{Density of States}

 Exploring the statistical mechanics of polymer systems requires considering the Liouville theorem, which is associated with the number of accessible microstates. In statistical mechanics, the phase space volume $\omega$ is pivotal since it allows us to determine the partition function and consequently compute other thermodynamic quantities.

 The Liouville volume of the $2n$-dimensional symplectic manifold is given as
\begin{eqnarray}\label{Vol-n}
\omega^{_n}=\frac{(-1)^{n(n-1)/2}}{n!}\,\underbrace{\omega\,\wedge
\,...\,\wedge\,\omega}_{n\,\,\mbox{times}}.
\end{eqnarray}
On a two dimensional symplectic manifold ${\mathbf M}$, the Liouville volume is equivalent to the closed nondegenerate symplectic $2$-form, as defined in \eqref{Vol-n}, and can be considered as a polymeric phase space with symplectic structure $\omega$. According to Darboux's theorem, $\omega$ can always be stated in canonical form, $\omega=d\mathbf q\wedge\,d\mathbf p\,$. The polymeric phase space is a canonical phase space whit a periodic momentum, leading to a nontrivial topology for the momentum part of the manifold ${{\mathbf M}}$. Polymerization, which is incorporated into Hamiltonian and modifies it, can be generalized in another equivalent way, a deformed commutation relation with the non-deformed standard Hamiltonian. Mathematically, these two approaches are linked together by the Darboux transformation. Darboux's theorem guarantees the existence of canonical coordinates satisfying commutative algebra on any symplectic manifold, implying that any noncommutative Poisson algebra can be transformed into a commutative one. However, the deformed commutation relation approach is more relevant in statistical mechanics because noncommutativity of the Phase space variables, which includes all the polymeric effects, solely affects the number of microstates via the deformed density of states.

Based on the relation \eqref{PT}, we consider the following transformation for canonical  variables $(\mathbf q, \mathbf p)$
\begin{equation}\label{Non-C}
(\mathbf q,\mathbf p)\rightarrow\,(q,p)=\bigg(\mathbf q\,\, ,\,\,\frac{2}{\beta}\sin\Big(\frac{
\beta \mathbf p}{2}\Big)\bigg),
\end{equation}
This is a noncanonical transformation that maps the modified Hamiltonian $\rm H_{\beta}(\mathbf q,\mathbf p)=\frac{1-\cos({\beta} \mathbf p)}{m{\beta}^2}+V(\mathbf q)$ from the polymeric canonical phase space to the standard Hamiltonian $\rm H_{\beta}( q, p)=\frac{p^2}{2m}+V(q)\,$  on the noncanonical phase space with polymeric variables (q,p), and the polymer momentum $p$ is bounded as $p\in[-\frac{2}{
\beta},+\frac{2}{\beta})$. To find the density of states, we need the Jacobian of the transformation \eqref{Non-C}. The Jacobian determinant for this transformation is
\begin{equation}\label{Jacobi}
J = \frac{\partial ({q},{p})}{\partial (\mathbf q,\mathbf p)}=\sqrt{1-\sin^2(\frac{
\beta \mathbf p}{2})}=\sqrt{1-\big(\frac{\beta}{2}\big)^2\big(\frac{2}{\beta}\big)^2\sin^2\big(\frac{
\beta \mathbf p}{2}\big)}=\sqrt{1-(\frac{
\beta p}{2})^2}.
\end{equation}
So, the polymer-deformed density of states is given by the Jacobian \eqref{Jacobi} as

\begin{eqnarray}\label{C-NC}
\int_{\mid \mathbf p\mid < \frac{\pi}{\beta}}(...)\,d\omega(\mathbf q,\mathbf p)\,\longrightarrow\int_{\mid p\mid < \frac{2}{\beta}}(...)\,\frac{d{\omega(q,p)}}{J(q,p)}
=\int_{\mid p\mid < \frac{2}{\beta}}(...)\,\frac{dq\wedge\,dp}{\sqrt{1-(\frac{\beta p}{2})^2}.}
\end{eqnarray}
where $d\omega(\mathbf q,\mathbf p)$ is the infinitesimal volume of the
2-dimensional canonical phase space and $d{\omega}(q,p)/J(q,p)$ is its noncanonical counterpart.
The new noncanonical phase space volume should be invariant under the time evolution of the system. In Hamiltonian formalism, time evolution for any function of the phase space $F(q,p)$ is given by the Poisson brackets, $\frac{dF}{dt}=\{F,{\rm{H}}\}$ (in the absence of explicit time dependence of $F$). So, the equations of motion would be derived as follows,
\begin{eqnarray}\label{EoM}
\dot{q}\;=\frac{dq}{dt}=\{q,{\rm{H_\beta}}\}\;=\frac{p}{m}\,\sqrt{1-\big(\frac{\beta p}{2}\big)^2},\hspace{1cm}
{\dot{p}}={\frac{dp}{dt}}=\{p,{\rm{H_\beta}}\}=-\frac{\partial V}{\partial q}\,\sqrt{1-\big(\frac{\beta p}{2}\big)^2}.
\end{eqnarray}
The modified equations of motion (\ref{EoM}) tend to the standard ones in the limit of $\beta\rightarrow\,0$.

The total volume of the phase space, encompassing both noncanonical and canonical coordinates, remains invariant under symplectic transformations. Integrating the Liouville volume, which represents the 2-form structure for a two-dimensional manifold, yields this total volume as

\begin{eqnarray}\label{TVolume}
\frac{1}{h}\mbox{Vol}(\omega^1)=\frac{1}{h}\int{\omega^1}=\frac{1}{h}\int_{
-\frac{\pi}{{\beta}}}^{+\frac{\pi}{{\beta}}}d\mathbf p\times\int_{\texttt{L}}d\mathbf q=\frac{1}{h}\int_{\texttt{L}}dq\times\int_{-\frac{2}{
\beta}}^{+\frac{2}{\beta}}\frac{dp}{\sqrt{1-(
\frac{\beta p}{2})^2}}=2\pi\Big(
\frac{\texttt{L}}{{h\beta}}\Big),
\end{eqnarray}

where we restricted ourselves to a finite one-dimensional spatial volume $\texttt{L}$.

Equation \eqref{TVolume} displays that, unlike its divergent classical counterpart, the phase space volume is finite. An interesting point to note here is that the spatial part of the phase space must be quantized with respect to the discreteness length parameter $\beta$. In fact, $\it{l}_{\rm poly}=\hbar{\beta}=\beta_{0}\hbar l_{_{\rm Pl}}$ in which $\beta_0={O}(1)$ is a dimensionless coefficient that must be determined by experiments \cite{QGE1,QGE2,QGE6}. Equation \eqref{TVolume} can be rewritten as $\mbox{Vol}(\omega^1)=\rm{nh},$ where $\rm n$ is a positive integer that counts $\hbar{\beta}$ that exist in $\texttt{L}$. Also, a maximal momentum is obtained as $p_{ max}\sim\,\frac{\hbar}{{\it{l}}_{ poly}}\sim {{\beta}}^{-1}$.
 To study thermodynamics of the polymeric n-dimensional systems, we need to consider a $2n$-dimensional phase space \eqref{Vol-n}. The homogenous transformation for the $2n$-dimensional phase space with canonical variables $\mathbf q^i=(\mathbf q^1,\mathbf q^2,...,\mathbf q^n)$ and $\mathbf p_i=(\mathbf p_1,\mathbf p_2,...,\mathbf p_n)$ is read
from the relation \eqref{Non-C} as
\begin{equation}\label{nNC-T}
(\mathbf q^i,\mathbf p_i)\rightarrow\,(q^i,p_i)=\Big(\mathbf q^i\,\, ,\,\, \frac{2}{
\beta}\sin(\frac{\beta \mathbf p_i}{2})\Big),
\end{equation}

where  $ q^i=(q^1,q^2,...,q^n)$ and $p_i=(p_1,p_2,...,p_n)$ are noncanonical corresponding variables. The $2$-form symplectic structure on the $2n$-dimensional symplectic manifold in the noncanonical chart becomes
\begin{eqnarray}\label{nc-v}
\omega=\sum_{i=1}^n\frac{dq^i\wedge dp_i}{\sqrt{1-(\frac{\beta p_i}{2})^2}}=\frac{dq^1\wedge dp_1}{\sqrt{1-(\frac{\beta p_1}{2})^2}}
+...+ \frac{dq^n\wedge dp_n}{\sqrt{1-(\frac{\beta p_n}{2})^2}},
\hspace{.7cm}
\end{eqnarray}

where the momenta are bounded as $p_i\in[-\frac{2}{\beta},+\frac{2}{\beta})$. The associated $2n$-form Liouville volume can be obtained by substituting \eqref{nc-v} into the definition \eqref{Vol-n} as

\begin{eqnarray}\label{L2nD}
\omega^n=\frac{dq^1\wedge dq^2\wedge...\wedge dq^n\wedge dp_1\wedge dp_2\wedge ... \wedge dp_n}{
\sqrt{\Big(1-(\frac{\beta p_1}{2})^2\Big)\Big(1-(\frac{
\beta p_2}{2})^2\Big)...\Big(1-(\frac{\beta p_n}{2})^2\Big)}},
\end{eqnarray}

So, the polymeric density of states, in noncanonical chart, can be easily deduced by
integrating over the Liouville volume \eqref{L2nD}  as
\begin{eqnarray}\label{nDOS}
\frac{1}{h^n}\mbox{Vol}(\omega^n)=\frac{1}{h^n}\int_{\mathbf M}\omega^n=\frac{1}{h^n}\int
d^nq\times\prod_{i=1}^n\int_{|p_i|<\frac{2}{\beta}}\frac{dp_i}{\sqrt{1-(\frac{\beta p_i}{2})^2}}\hspace{6cm}\\\nonumber=\frac{1}{h^n}\int
d^nq\underbrace{\int_{-\frac{2}{\beta}}^{+\frac{2}{\beta}}\int_{-\frac{2}{\beta}}^{+\frac{2}{\beta}}\,...\,\int_{-\frac{2}{\beta}}^{+\frac{2}{\beta}}}_{n\,\,\mbox{times}}\frac{
d{p_1}\,d{p_2}\,...\,d{p_n}}{\sqrt{\Big(1-(\frac{\beta
p_1}{2})^2\Big)\Big(1-(\frac{\beta p_2}{2})^2\Big)\,...\,\Big
(1-(\frac{\beta p_n}{2})^2\Big)}}.
\end{eqnarray}

This relation gives the deformed density of states.

\subsection{Partition Function}
Statistical mechanics links macrostates and  microstates. A system's thermodynamical features can be obtained from its partition function, which sums over all accessible microstates. Using the deformed density of state (\ref{nDOS}) and replacing the Hamiltonian, the  partition function for the polymeric phase space is given by
\begin{eqnarray}\label{Dar-PF-D}
{\mathcal Z}_1({\beta};T)=\frac{1}{h^n}\int d^nq\prod_{i=1}^n
\int_{_{-\frac{2}{\beta}}}^{^{+\frac{2}{\beta}}}\,\frac{dp_i}{\sqrt{1-(\frac{\beta p_i}{2})^2}}\exp\Big[
-\frac{\mbox H(q^i,p_i)}{T}\Big].
\end{eqnarray}

 Having the partition function (\ref{Dar-PF-D}), one can study the thermodynamics of any physical system in the semiclassical polymer framework. We use the polymeric partition function (\ref{Dar-PF-D}) to study the thermodynamics of a system of $n$-Dimensional Harmonic Oscillators in the next section. Our focus in this framework is on the issue of dimensional reduction\footnote{Throughout this article, we work in units $c=\hbar=k_{_B}=1$, where $c$ and $k_B$ are the speed of light in vacuum and the Boltzmann constant, respectively}.

\section{Thermodynamics}
\subsection{Thermodynamical Dimensional Reduction of a System of n-Dimensional Harmonic Oscillators in Polymeric framework}

In this section, we examine a system of n-dimensional Harmonic Oscillators in the polymer framework. The corresponding Hamiltonian function for a single harmonic oscillator is $\mbox H=\frac{p^2}{2m}+\frac{1}{2}m\omega^2q^2$ and by substituting it in Eq. \eqref{Dar-PF-D},
the corresponding non-canonical polymeric partition function in semi-classical regime can be derived as follows

\begin{eqnarray}\label{PF-D}
{\mathcal Z}_1(T)=\frac{2\pi^{n/2}}{h^n\Gamma \big(\frac{n}{2}\big)}\int_{q_{min}}^{\infty}
\exp\left[-\frac{m{\omega^2}q^2}{2T}\right]q^{n-1}dq\prod_{i=1}^n\Bigg(
\int_{-\frac{2 }{\beta }}^{\frac{2 }{\beta }}\exp\left[-\frac{
{p_i}^2}{2mT}\right]\frac{dp_i}{\sqrt{1-(\frac{\beta p_i}{2})^2}}\Bigg)\nonumber\\
=\frac{(2\pi)^{\frac{3n}{2}}}{h^n \left(\frac{m w^2}{T}\right)^{\frac{n}{2}}\beta^n}
\Bigg(\exp\left[-\frac{1}{mT{\beta}^2}\right]\,I_{0}\left[
\frac{1}{mT{\beta}^2}\right]\Bigg)^n,
\end{eqnarray}

where $I_0$ denotes the zero rank of the first kind of the modified Bessel function. Rewriting this partition function versus the thermal de Broglie wavelength $\lambda=\frac{h}{\sqrt{2\pi m T}}$, gives

\begin{eqnarray}\label{PF1-IG}
{\mathcal Z}_{1}(\lambda)=\frac{(2\pi )^{\frac{3n}{2}} \ell^{2n}}{\lambda^n\lambda _{_{\rm{Pl}}}^n}\Bigg(\exp\left[-\frac{\lambda^2}{\lambda_{_{\rm{Pl}}}^2}\right]I_0
\left[\frac{\lambda^2}{\lambda_{_{\rm{Pl}}}^2}\right]\Bigg)^{n}.
\end{eqnarray}

In this relation, $\ell=\sqrt{\frac{1}{m\omega}}$ is the characteristic length of the harmonic oscillator system and ${\lambda_{_{\rm Pl}}}$ is the thermal de Broglie wavelength in the framework of the polymer quantization and it is of the order of Planck scale thermal de Broglie wavelength as follows
\begin{equation}\label{lambda-GUP}
\lambda_{_{\rm Pl}}=\sqrt {2\pi}\beta=\sqrt{2\pi}\beta_0\rm l_{_{\rm Pl}}=\frac{h\beta_0}{\sqrt{
2\pi{m_{_{\rm Pl}}}T_{_{\rm Pl}}}}
\end{equation}
where $m_{_{\rm Pl}}$ is the Planck mass (equal to the Planck temperature in the units adopted in this paper).\\
Expanding the partition function, Eq. (\ref{PF1-IG}), for high and low temperature regimes is given by (See Appendix A for details.)
\begin{equation}\label{zp0-Exp}
{\mathcal Z}_1[\lambda]\approx
\left\{
  \begin{array}{ll}
    \begin{gathered}
    (2\pi)^n\frac{ \ell^{2n}}{\lambda^{2n}}=\Big(\frac{T}{\hbar \omega}\Big)^n\end{gathered} & \hspace{.5cm}\lambda
    \gg\lambda_{_{\rm Pl}},\\\\
   \begin{gathered} (2\pi )^n\frac{\ell^{2n}}{\lambda^n\lambda _{_{\rm{Pl}}}^n} \end{gathered}&
    \hspace{.5cm}\lambda\sim\lambda_{_{\rm Pl}}\,,
  \end{array}
\right.
\end{equation}

where $\mathcal{Z}=\big(\frac{T}{\hbar \omega}\big)^n$ is the non-deformed classical partition function for an $n$-dimensional single-harmonic oscillator. As we see from the above expansion, at high temperatures (or when the thermal de Broglie wavelength approaches the Planck thermal de Broglie wavelength, $\lambda\sim\lambda_{\rm Pl}\,(T\sim{T}_{\rm Pl})$), the system's degrees of freedom reduce from $2n$ to $n$ (${\lambda^{2n}}\longrightarrow\,{\lambda^n}$). This effect becomes negligible at low temperatures ($\lambda\gg\lambda_{\rm Pl}\,(T\ll{T}_{\rm Pl})$), recovering the standard harmonic oscillator partition function.

For the total partition function in this picture, we have
\begin{eqnarray}\label{TPF-IG}
{\mathcal Z}_{\mathcal{N}}(\lambda)=\left(\frac{(2\pi )^\frac{3n}{2} \ell^{2n}}{\lambda^n\lambda _{_{\rm{Pl}}}^n}\right)^\mathcal{N}\Bigg(\exp\left[-\frac{\lambda^2}{\lambda_{_{\rm{Pl}}}^2}\right]I_0
\left[\frac{\lambda^2}{\lambda_{_{\rm{Pl}}}^2}\right]\Bigg)^{n\,\mathcal{N}}\,,
\end{eqnarray}

where the Gibbs factor, $\frac{1}{\mathcal{N}!}$ is omitted as it is assumed that the oscillators are localized ~\cite{SMB1,SMB2}.
By the relation (\ref{TPF-IG}), the Helmholtz free energy $F=-T\ln({\mathcal Z}_{\mathcal{N}}(\lambda))$ can be derived as

\begin{eqnarray}\label{Helmholtz-p}
F=-\mathcal{N}T\ln\Bigg[\frac{(2\pi )^\frac{3n}{2}  \ell^{2n}}{\lambda^n\lambda _{_{\rm{Pl}}}^n}\Bigg(\exp\left[-\frac{\lambda^2}{\lambda_{_{\rm{Pl}}}^2}\right]I_0
\left[\frac{\lambda^2}{\lambda_{_{\rm{Pl}}}^2}\right]\Bigg)^{n}\Bigg].
\end{eqnarray}

\subsubsection{Internal Energy}

The corresponding internal energy $U=-T^2\big(\frac{\partial}{\partial T}(
\frac{F}{T})\big)$ for such a system is as follows
\begin{eqnarray}\label{energy-p}
U=n\mathcal{N}T\Bigg(\frac{1}{2}+\frac{\lambda^2}{\lambda_{_{\rm{Pl}}}^2}\bigg(1-\frac{I_1\big[
\frac{\lambda^2}{\lambda_{_{\rm{Pl}}}^2}\big]}{I_0\big[\frac{\lambda^2}{\lambda_{_{\rm{Pl}}}^2}\big]}
\bigg)\Bigg)\,.
\end{eqnarray}

 The internal energy for a usual n-dimensional harmonic oscillator system linearly depends on the temperature through the well-known relation $U=n\mathcal{N}T$ and any high energy scale is accessible for the system just by sufficiently increasing the temperature. The corresponding relation for the polymeric n-dimensional harmonic oscillator is given by the relation (\ref{energy-p}) which leads to $U=\frac{1}{2}n\mathcal{N}T$ at a high temperature limit

\begin{equation}\label{energy-expand}
U=
\left\{
  \begin{array}{ll}
    n\mathcal{N}T & \hspace{.5cm}\lambda
    \gg\lambda_{_{\rm Pl}},\\\\
  \,\frac{1}{2}n\mathcal{N}T &
    \hspace{.5cm}\lambda\sim\lambda_{_{\rm Pl}}\,.
  \end{array}
\right.
\end{equation}

\begin{figure}[ht]
  \centering
  \begin{subfigure}[b]{0.35\textwidth}
    \includegraphics[width=3in]{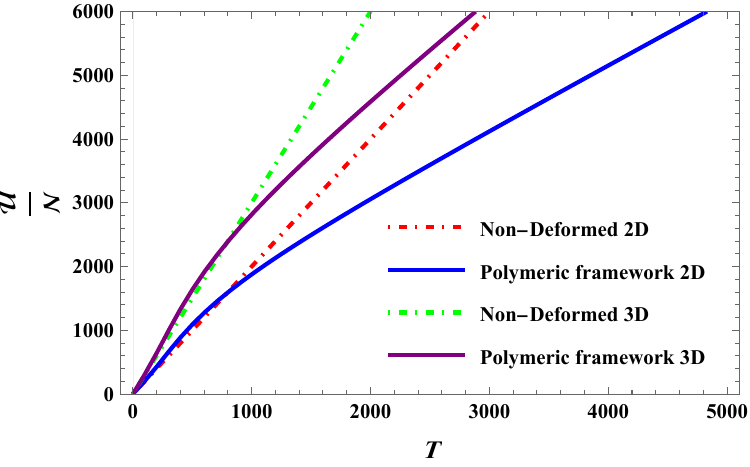}
    \caption{}
    \label{fig:1sub1}
  \end{subfigure}
  \hfill
  \begin{subfigure}[b]{0.48\textwidth}
    \includegraphics[width=3in]{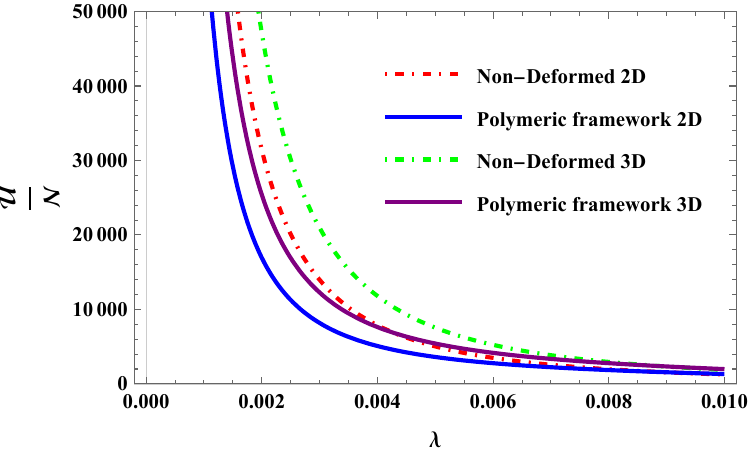}
    \caption{}
    \label{fig:1sub2}
  \end{subfigure}
  \caption{ The internal energy in terms of temperature (left), and the internal energy in terms of thermal de Broglie wavelength (right). The figures are plotted in units of $ \mathcal N=5, m=100, \omega=10, \beta = 0.004$.}
  \label{fig:1}
\end{figure}

Figure (\ref{fig:1sub1}) illustrates the internal energy of 2D and 3D harmonic oscillators as a function of temperature $T$. It is observed that at high temperatures (Planck temperature), the system's internal energy deviates significantly from its classical state. However, as the temperature decreases, this deviation gradually diminishes until it eventually aligns with the classical state. Figure (\ref{fig:1sub2}) displays the internal energy of 2D and 3D harmonic oscillators in terms of thermal de Broglie wavelength $\lambda$. It is evident that at extremely small wavelengths (polymeric length effects), the deviation from the classical state intensifies. Conversely, as the thermal wavelength increases, the behavior of the internal energy approaches its classical state.

\begin{figure}[ht]
  \centering
  \begin{subfigure}[b]{0.45\textwidth}
    \includegraphics[width=3in]{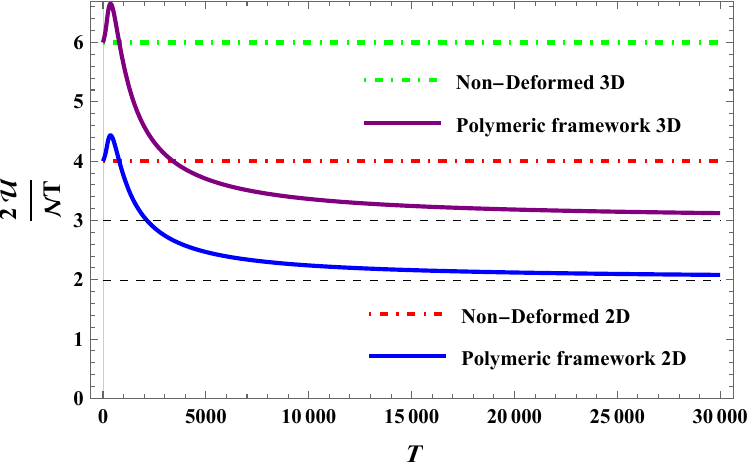}
    \caption{}
    \label{fig:2sub1}
  \end{subfigure}
  \hfill
  \begin{subfigure}[b]{0.45\textwidth}
    \includegraphics[width=3in]{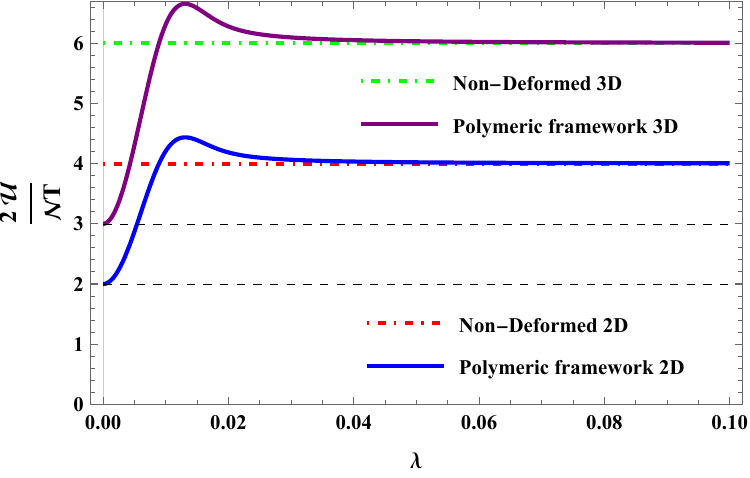}
    \caption{}
    \label{fig:2sub2}
  \end{subfigure}
  \caption{The number of degrees of freedom in terms of temperature (left), and the number of degrees of freedom in terms of thermal de Broglie wavelengththe (right). The figures are plotted in units of $ \mathcal N=5, m=100, \omega=10, \beta = 0.004$.}
  \label{fig:2}
\end{figure}

Figure (\ref{fig:2sub1}) illustrates the number of degrees of freedom of 2D and 3D harmonic oscillators as a function of temperature $T$. Based on the classical equipartition theorem of energy, the number of degrees of freedom of a system can be determined by the formula $\frac{(U/N)}{(T/2)}$. It is observed that in the Polymer framework, this value is temperature-dependent and decreases from $6$ to $3$ for a 3D harmonic oscillator system and from $4$ to $2$ for a 2D harmonic oscillator system at extremely high temperatures. Figure (\ref{fig:2sub2}) shows the number of degrees of freedom in terms of thermal de Broglie wavelength $\lambda$. At extremely small wavelengths, the system's degrees of freedom are reduced to two and three for 2D and 3D harmonic oscillators respectively.

\subsubsection{Specific Heat}

The specific heat in polymeric picture, which can be obtained from the internal energy (\ref{energy-p}) via the relation $C_{_V}=\big(\frac{\partial U}{\partial T}
\big)$, is given by

\begin{eqnarray}\label{SHeat-p}
C_{_V}=\frac{1}{2}n\mathcal{N}\Bigg(1+\frac{\lambda^4}{\lambda_{_{\rm Pl}}^4}\bigg(1+\frac{I_2\big[\frac{\lambda^2}{\lambda_{_{\rm Pl}}^2}\big]}{I_0\big[\frac{
\lambda^2}{\lambda_{_{\rm Pl}}^2}\big]}-2\Big(\frac{I_1\big[\frac{\lambda^2}{\lambda_{_{\rm Pl}}^2}\big]}{
I_0\big[\frac{\lambda^2}{\lambda_{_{\rm Pl}}^2}\big]}\Big)^2\bigg)\Bigg)\,.\hspace{1cm}
\end{eqnarray}
 Expansion of the specific
heat \eqref{SHeat-p} for the low and high temperatures regimes gives
\begin{equation}\label{Sp-Heat-Exp}
C_{_V}=
\left\{
  \begin{array}{ll}
    n\mathcal{N} & \hspace{.5cm}\lambda
    \gg\lambda_{_{\rm Pl}},\\\\
    \frac{1}{2}n\mathcal{N} &
    \hspace{.5cm}\lambda\sim\lambda_{_{\rm Pl}}\,,
  \end{array}
\right.
\end{equation}

\begin{figure}[ht]
  \centering
  \begin{subfigure}[b]{0.35\textwidth}
    \includegraphics[width=3in]{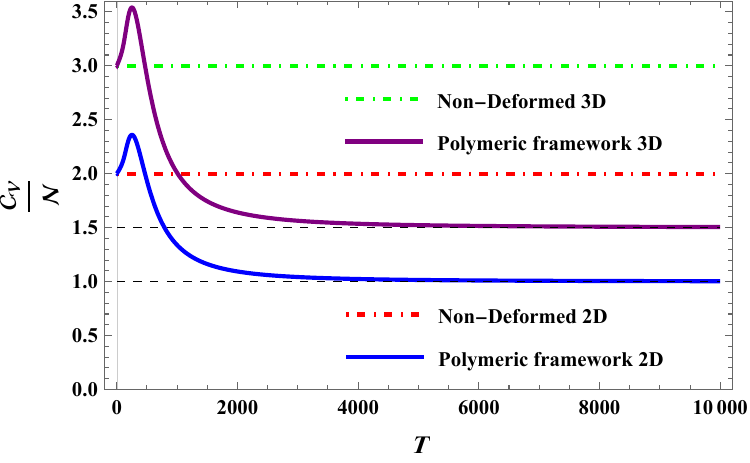}
    \caption{}
    \label{fig:3sub1}
  \end{subfigure}
  \hfill
  \begin{subfigure}[b]{0.48\textwidth}
    \includegraphics[width=3in]{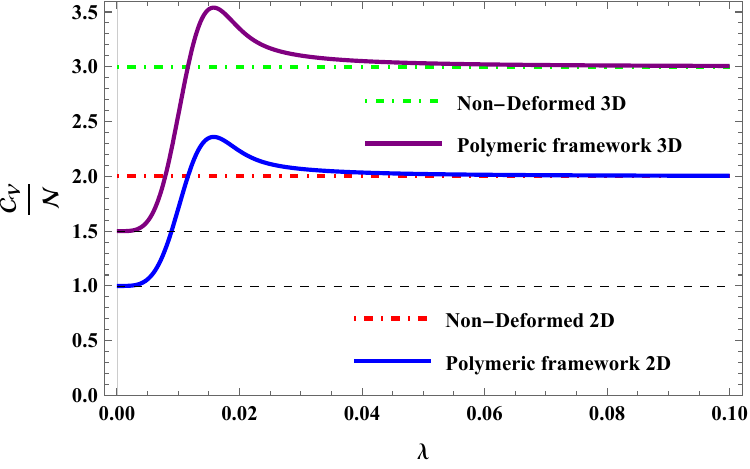}
    \caption{}
    \label{fig:3sub2}
  \end{subfigure}
  \caption{The specific heat in terms of temperature (left), and the specific heat in terms of thermal de Broglie wavelengththe (right). The figures are plotted in units of $ \mathcal N=5, m=100, \omega=10, \beta = 0.004$.}
  \label{fig:3}
\end{figure}

As we see, this quantity shows a decreasing behavior in high temperature and eventually tends to 1D and 1.5D for 2D and 3D harmonic oscillators respectively. (see figure \ref{fig:3}).

Figur (\ref{fig:3sub1}) illustrates the specific heat of 2D and 3D harmonic oscillators (HO) as a function of temperature $T$. The system's specific heat is temperature-dependent for the Polymeric framework and asymptotically approaches $1$ (for 2D-HO) and $1.5$ (for 3D-HO) at the very high temperature regime, which signals the effective reduction of the degrees of freedom from $4$ to $2$ (2D-HO) and $6$ to $3$ (3D-HO) in this setup. It is also clear from the figure that the specific heat is bounded as $1\leq\frac{C_{_V}}{\mathcal N}\leq 2$ (2D-HO) and $1.5 \leq\frac{C_{_V}}{\mathcal N}\leq 3$ (3D-HO) in the Polymeric framework. Figure (\ref{fig:3sub2}) shows the specific heat in terms of thermal de Broglie wavelength. It is observed that at very small wavelengths, the deviation from the classical state intensifies and finally reaches the value of $1$ and $\frac{3}{2}$ for 2D and 3D harmonic oscillators respectively.

\subsubsection{Entropy}

Finally, from the Helmholtz free energy (\ref{Helmholtz-p}), the entropy of the polymeric harmonic oscillator, $S=-\left(\frac{\partial F}{\partial T} \right)$, can be derived as follows

\begin{eqnarray}\label{entropy-p}
S=\mathcal{N}\left(\frac{1}{2} n\Bigg(1+\frac{2\lambda^2}{\lambda_{_{\rm{Pl}}}^2}\bigg(1-\frac{I_1\big[
\frac{\lambda^2}{\lambda_{_{\rm Pl}}^2}\big]}{I_0\big[\frac{\lambda^2}{\lambda_{_{\rm Pl}}^2}\big]}\bigg)\Bigg)+\ln
\left[\frac{(2\pi )^\frac{3n}{2}  \ell^{2n}}{\lambda^n\lambda _{_{\rm{Pl}}}^n}\Bigg(\exp\left[-\frac{\lambda^2}{\lambda_{_{\rm{Pl}}}^2}\right]I_0
\left[\frac{\lambda^2}{\lambda_{_{\rm{Pl}}}^2}\right]\Bigg)^{n}
\right]\right)\,.
\end{eqnarray}

\begin{figure}[ht]
  \centering
  \begin{subfigure}[b]{0.35\textwidth}
    \includegraphics[width=3in]{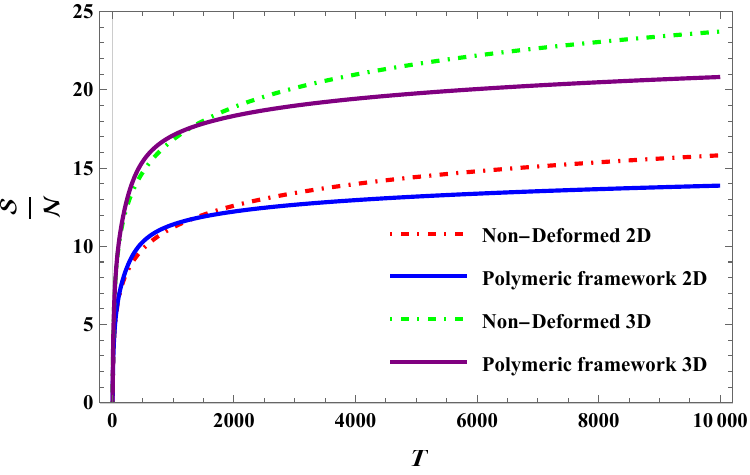}
    \caption{}
    \label{fig:4sub1}
  \end{subfigure}
  \hfill
  \begin{subfigure}[b]{0.48\textwidth}
    \includegraphics[width=3in]{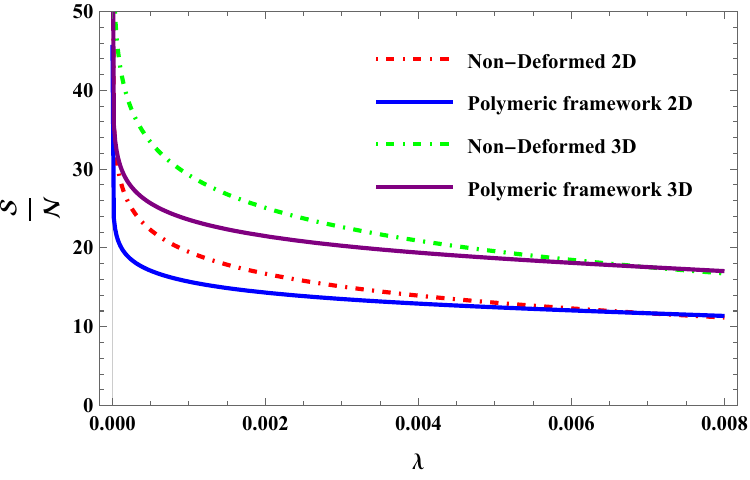}
    \caption{}
    \label{fig:4sub2}
  \end{subfigure}
   \caption{ The entropy in terms of temperature (left), and the entropy in terms of thermal de Broglie wavelengththe (right). The figures are plotted in units of $ \mathcal N=5, m=100, \omega=10, \beta = 0.004$.}
  \label{fig:4}
\end{figure}

As we see in Figure \ref{fig:4}, the thermodynamic outcome is that the entropy decreases in the polymer framework since entropy is determined directly from the number of microstates, which is reduced in this framework due to the increase in the fundamental cell volume.

 Figure (\ref{fig:4sub1}) illustrates the entropy of 2D and 3D harmonic oscillators as a function of temperature $T$. It is evident that, in the Polymer framework, the entropy increases at a slower rate than in the standard non-deformed case. This is due to the reduction in the number of accessible microstates in the high temperature regime, caused by the effects of quantum gravity. Figure (\ref{fig:4sub2}) illustrates the entropy in relation to the thermal de Broglie wavelength $\lambda$. It is observed that at extremely small wavelengths, the entropy demonstrates a slower growth rate compared to the classical non-deformed scenario.\\

 Before summarizing our findings, we note that in Ref.~\cite{Kajuri:2014kva}, the author has shown that the path integral formulation of a polymer
quantum oscillator acquires an extra integer summation over the windings/images (see for instance Eq. (23) of this reference). This extra integer summation is expected in systems with periodic boundaries and should appear in the partition function as well. In our case we have not considered this extra
summation since we are dealing mainly with the high temperature regime that the effect of dimensional reduction is dominant. Actually in the high temperature regime the case of $l=0$ sector typically dominates in analysis of Ref.~\cite{Kajuri:2014kva}. At low temperature, the situation is different and other sectors with $l\neq 0$ images should be considered. In the language of our setup, as has been shown in subsection III-A, the dimensional reduction effect becomes negligible at low temperatures (that is, $\lambda\gg\lambda_{\rm Pl}\,(T\ll{T}_{\rm Pl})$), recovering the standard harmonic oscillator partition function without dimensional reduction. A detailed analysis of the low temperature limit in the line of the mentioned reference may be considered in a separate study.

\section{Summary and Conclusions}
In this study, we explored the statistical mechanics and the thermodynamic behavior of an $n$-dimensional system of harmonic oscillators within the polymer quantization framework. We analyzed the results for two- and three-dimensional harmonic oscillators both analytically and numerically. By employing a non-canonical transformation for the phase space variables, we derived the deformed density of states and the polymeric-deformed partition function analytically. We also calculated the corresponding thermodynamic quantities, including internal energy, specific heat, and entropy, in the semi-classical regime. A significant result is a thermodynamic dimensional reduction in high-temperature regime, where the thermal de Broglie wavelength approaches the scale of the polymer. This dimensional reduction which occurs continuously, effectively reduces the number of thermodynamic degrees of freedom of the system. Our analysis shows that polymer effect, applied to the density of states, reduces the number of accessible microstates. This feature results in a limited specific heat and entropy, significantly altering the usual thermodynamic behavior. In the low-energy limit, these deviations disappear and ordinary classical results are restored. While this research is phenomenological and semiclassical, it offers beneficial insights into the thermodynamic consequences of minimal length and maximal momentum effects.
In brief, the polymer representation of quantum mechanics offers a valuable framework for incorporating phenomenological quantum gravity effects into statistical mechanics. It also unlocks methods for gaining a deeper understanding of dimensionality and spacetime microstructure from a thermodynamic point of view.

\begin{acknowledgments}
The authors sincerely thank the referees for their constructive and insightful comments.
\end{acknowledgments}
\appendix
\renewcommand{\theequation}{A-\arabic{equation}}
\setcounter{equation}{0}

\section{Expansion of the Partition Function}
Consider the coefficient in the partition function in Eq.~\eqref{PF1-IG} as
\begin{equation}\label{FL}
F(\lambda) = \Bigg(\exp\left[-\frac{\lambda^2}{\lambda_{_{\rm{Pl}}}^2}\right]I_0
\left[\frac{\lambda^2}{\lambda_{_{\rm{Pl}}}^2}\right]\Bigg)^{n},
\end{equation}
where $\lambda_{_{\rm{Pl}}} \ll 1$ (very small). We define, $x\equiv\frac{\lambda^2}{\lambda_{_{\rm{Pl}}}^2}$.
\section*{(A) $\lambda\gg\lambda_{_{\rm{Pl}}}$}
 In the regime where $\lambda \gg \lambda_{_{\rm{Pl}}}$ (i.e.\ $x \gg 1$), for large $x$, the Bessel function behaves as
\begin{equation}
I_0(x)\sim \frac{e^{x}}{\sqrt{2\pi x}}\left(1+\frac{1}{8x}+\frac{9}{2!(8x)^2}+\frac{225}{3!(8x)^3}+O(x^{-4})\right).
\end{equation}

Multiplying by $e^{-x}$ gives

\begin{equation}
e^{-x}I_0(x)=(2\pi x)^{-1/2}\,(1+\varepsilon(x)),
\qquad \varepsilon(x)=\frac{1}{8x}+\frac{9}{128x^2}+O(x^{-3}).
\end{equation}

Raising to the power $n$ and expanding up to $O(x^{-2})$ yields

\begin{align}
\bigl(e^{-x}I_0(x)\bigr)^n &= (2\pi)^{-n/2} x^{-n/2}\,(1+\varepsilon(x))^n \\
&= (2\pi)^{-n/2} x^{-n/2}\left(1+n\varepsilon(x)+\frac{n(n-1)}{2}\varepsilon(x)^2+O(\varepsilon^3)\right).\notag
\end{align}

Keeping terms up to $O(x^{-2})$ we substitute $\varepsilon(x)$ and compute
\begin{align}
n\varepsilon(x) &= \frac{n}{8x}+\frac{9n}{128x^2}+O(x^{-3}),\\[6pt]
\frac{n(n-1)}{2}\varepsilon(x)^2 &= \frac{n(n-1)}{2}\cdot\frac{1}{64x^2}+O(x^{-3})
= \frac{n(n-1)}{128x^2}+O(x^{-3}).
\end{align}

Thus the expansion is

\begin{equation}\label{exp_power}
\bigl(e^{-x}I_0(x)\bigr)^n \sim (2\pi)^{-n/2} x^{-n/2}\left(1+\frac{n}{8x}+\frac{n^2+8n}{128x^2}+O(x^{-3})\right).
\end{equation}

Returning to \(\lambda\) using $x=\frac{\lambda^2}{\lambda_{_{\rm{Pl}}}^2}$ (so $x^{-1}=\frac{\lambda_{_{\rm{Pl}}}^2}{\lambda^2}$ and $x^{-n/2}=\frac{\lambda_{_{\rm{Pl}}}^n}{\lambda^n}$) gives

\begin{equation}
F(\lambda) = \Bigg(\exp\left[-\frac{\lambda^2}{\lambda_{_{\rm{Pl}}}^2}\right]I_0
\left[\frac{\lambda^2}{\lambda_{_{\rm{Pl}}}^2}\right]\Bigg)^{n} \sim (2\pi)^{-n/2}\left(\frac{\lambda_{_{\rm{Pl}}}}{\lambda}\right)^{\!n}
\left(1+\frac{n}{8}\frac{\lambda_{_{\rm{Pl}}}^2}{\lambda^2}+\frac{n^2+8n}{128}
\frac{\lambda_{_{\rm{Pl}}}^4}{\lambda^4}+O(\lambda_{_{\rm{Pl}}}^6)\right),
\end{equation}
or
\begin{equation}
F(\lambda) \sim (2\pi)^{-n/2}\Big(\frac{\lambda_{_{\rm{Pl}}}}{\lambda}\Big)^n
\Big(1+O((\lambda_{_{\rm{Pl}}}/\lambda)^2)\Big).
\end{equation}

\section*{(B) $\lambda\sim\lambda_{_{\rm{Pl}}}$}
In the regime where $\lambda$ is \emph{close} to $\lambda_{_{\rm{Pl}}}$ (i.e.\ $\lambda\to\lambda_{_{\rm{Pl}}}$), we define $\varepsilon=x-1$, so that $\varepsilon\ll 1$ when $\lambda$ is near $\lambda_{_{\rm{Pl}}}$.

We will use the Taylor expansion of $I_0$ about $x=1$. The needed derivatives can be written in terms of $I_0$ and $I_1$(using identities for $I_\nu$ derivatives),

\begin{equation}
I_0'(x)=I_1(x),\,\,\,\,\, I_0''(x)=I_0(x)-\frac{1}{x}I_1(x).
\end{equation}

Up to second order in $\varepsilon$,
\begin{equation}
I_0(1+\varepsilon)=I_0(1)+I_1(1)\,\varepsilon+\tfrac{1}{2}I_0''(1)\,\varepsilon^2+O(\varepsilon^3).
\end{equation}
and
\begin{equation}
e^{-x}=e^{-1}e^{-\varepsilon}=e^{-1}\Big(1-\varepsilon+\tfrac{1}{2}\varepsilon^2+O(\varepsilon^3)\Big).
\end{equation}

Multiplying the two above expansions and collect terms up to order $\varepsilon^2$, we have
\begin{equation}
e^{-x}I_0(x)=a_0 + a_1\varepsilon + a_2\varepsilon^2 + O(\varepsilon^3),
\end{equation}
where
\begin{equation}
\begin{aligned}
a_0 &= e^{-1}I_0(1),\\[4pt]
a_1 &= e^{-1}\bigl(I_1(1)-I_0(1)\bigr),\\[4pt]
a_2 &= e^{-1}\Big(\tfrac{1}{2}I_0(1)-I_1(1)+\tfrac{1}{2}I_0''(1)\Big).
\end{aligned}
\end{equation}

Raising to the power $n$ and expanding up to $O(\varepsilon^{3})$ yields

\begin{equation}
\begin{aligned}
F(\lambda)=\bigl(e^{-x}I_0(x)\bigr)^n \sim a_0^{\,n}\Big(1+\tfrac{a_1}{a_0}\varepsilon+\tfrac{a_2}{a_0}\varepsilon^2
+O(\varepsilon^3)\Big)^n\sim a_0^{\,n}\Bigg[
1 &+ n\frac{a_1}{a_0}\varepsilon \\
&+ \left( n\frac{a_2}{a_0} + \frac{n(n-1)}{2}\Big(\frac{a_1}{a_0}\Big)^2 \right) \varepsilon^2
+ O(\varepsilon^3)
\Bigg].
\end{aligned}
\end{equation}

or
\begin{equation}
F(\lambda)\sim a_0^{\,n}\Bigg[1+ n\frac{a_1}{a_0}\varepsilon +O(\varepsilon^2)\Bigg].
\end{equation}

At $\lambda=\lambda_{_{\rm{Pl}}}$ (equivalently $\varepsilon=0$), we have

\begin{equation}
F(\lambda_{_{\rm{Pl}}})=a_0^{\,n}=\Bigl(\frac{I_0(1)}{e}\Bigr)^{\!n}\approx(0.4657596075936405)^{n}\approx\left(\frac{1}{\sqrt{2\pi}}\right)^n.
\end{equation}

\end{document}